\documentclass[a4paper,fleqn,usenatbib]{mnras}

\usepackage{newtxtext,newtxmath}
\usepackage[T1]{fontenc}
\usepackage{ae,aecompl}

\usepackage{graphicx}	% Including figure files
\usepackage{amsmath}	% Advanced maths commands
\newcommand{\diffp}[2]{\frac{\partial #1}{\partial #2}}
\renewcommand{\vec}[1]{\boldsymbol{#1}}

\title[Magnetohydrodynamics and \emph{Ulysses} observations]{Flux conservation, radial scalings, Mach numbers, and critical distances in the solar wind: magnetohydrodynamics and \emph{Ulysses} observations}

\author[D.~Verscharen, S.~D.~Bale and M.~Velli]{
Daniel Verscharen,$^{1,2}$\thanks{E-mail: d.verscharen@ucl.ac.uk}
Stuart D.~Bale$^{3,4,5,6}$ and
Marco Velli$^{7}$
\\
$^{1}$Mullard Space Science Laboratory, University College London, Holmbury House, Dorking, RH5\,6NT, UK\\ 
$^{2}$Space Science Center, University of New Hampshire, Durham, NH 03824, USA\\
$^{3}$Physics Department, University of California, Berkeley, CA 94720-7300, USA\\
$^{4}$Space Sciences Laboratory, University of California, Berkeley, CA 94720-7450, USA\\
$^{5}$The Blackett Laboratory, Imperial College London, London, SW7\,2AZ, UK\\
$^{6}$School of Physics and Astronomy, Queen Mary University of London, London, E1\,4NS, UK\\
$^{7}$Department of Earth, Planetary, and Space Sciences, University of California Los Angeles, Los Angeles, CA 90095, USA
}

\date{Accepted 2021 July 13. Received 2021 July 7; in original form 2021 April 8}

\pubyear{2021}

\begin{document}
\label{firstpage}
\pagerange{\pageref{firstpage}--\pageref{lastpage}}
\maketitle

\begin{abstract}
One of the key challenges in solar and heliospheric physics is to understand the acceleration of the solar wind. As a super-sonic, super-Alfv\'enic plasma flow, the solar wind carries mass, momentum, energy, and angular momentum from the Sun into interplanetary space. We present a framework  based on two-fluid magnetohydrodynamics to estimate the flux of these quantities based on spacecraft data independent of the heliocentric distance of the location of measurement. Applying this method to the \emph{Ulysses} data set allows us to study the dependence of these fluxes on heliolatitude and solar cycle. The use of scaling laws provides us with the heliolatitudinal dependence and the solar-cycle dependence of the scaled Alfv\'enic and sonic Mach numbers as well as the Alfv\'en and sonic critical radii. Moreover, we estimate the distance at which the local thermal pressure and the local energy density in the magnetic field balance. These results serve as predictions for observations with \emph{Parker Solar Probe}, which currently explores the very inner heliosphere, and \emph{Solar Orbiter},  which will measure the solar wind outside the plane of the ecliptic in the inner heliosphere during the course of the mission.
\end{abstract}

\begin{keywords}
solar wind -- Sun: heliosphere -- magnetohydrodynamics -- plasmas -- methods: data analysis
\end{keywords}

%%%%%%%%%%%%%%%%%%%%%%%%%%%%%%%%%%%%%%%%%%%%%%%%%%

%%%%%%%%%%%%%%%%% BODY OF PAPER %%%%%%%%%%%%%%%%%%

\section{Introduction}

The Sun, like most other stars, continuously emits a magnetized plasma in the form of the solar wind \citep{verscharen19}. This super-sonic and super-Alfv\'enic flow fills the interplanetary space and removes mass, momentum, energy, and angular momentum from the Sun. The acceleration mechanisms of the solar wind remain poorly understood and pose one of the greatest science questions in the field of solar and heliospheric physics. Since the early time of the space age, starting in the early 1960s, a fleet of spacecraft have measured the properties of the solar wind at different locations in the heliosphere. The \emph{Ulysses} mission \citep{wenzel92,balogh94,marsden01}, in operation from 1990 until 2009, plays a special role amongst them  due to its unique orbit that led the spacecraft above the Sun's poles, enabling studies of the solar-wind parameters as functions of heliolatitude. These studies are of great importance to the question of the solar-wind acceleration, since they enable the separation of different solar-wind source regions and their relationships to the  heliolatitude-dependent magnetic-field structure in the corona \citep{neugebauer99}. Before \emph{Ulysses}, all solar-wind missions were restricted to quasi-equatorial orbits. These measurements could only be used to explore heliomagnetic latitudes up to $\pm 25^{\circ}$ \citep{bruno86}, exploiting the tilt between the Sun's magnetic dipole axis and its rotation axis.
The knowledge about the dependence of solar-wind parameters on radial distance and heliolatitude helps constrain models for our understanding of the acceleration of the solar wind. For example, comparisons of in situ mass-flux measurements with coronagraph observations suggest that the solar wind requires an additional deposition of energy to the contribution from thermal conduction alone \citep{munro77} as assumed in the classic \citet{parker58} model of the solar wind. Moreover, measurements of the plasma's mass flux can be linked to photospheric measurements of the magnetic field, which provide us with insight into the location and the magnetic nature of the solar-wind heating processes \citep{wang10}. \emph{Ulysses} data confirm this need for additional energy deposition also in polar wind \citep{barnes95}.

\emph{Ulysses} observations corroborated the bimodal structure of the solar wind during solar minimum \citep{mccomas98,mccomas98b}: near the Sun's equator at heliolatitudes below approximately $\pm 20^{\circ}$, the wind is variable and slow (radial flow speeds $\lesssim 400\,\mathrm{km\,s}^{-1}$); in polar regions, the wind is steadier and fast (radial flow speeds $\gtrsim 700\,\mathrm{km\,s}^{-1}$). During solar maximum, this bimodality vanishes almost completely, and the solar wind exhibits large variations in its plasma and field parameters \citep{mccomas00}. The measurements have been effectively visualized in polar plots, in which the polar angle indicates the heliolatitude and the distance from the origin indicates the solar-wind parameter \citep[e.g., speed and density, see][Plate 1]{mccomas00}. These comprehensive studies of \emph{Ulysses} data also provide us with fit results  for solar-wind parameters depending on heliocentric distance and heliolatitude \citep[see also][]{ebert09}.

Although the Sun's mass and energy loss due to the solar wind are insignificant throughout the Sun's life cycle, the loss of angular momentum carried away by the solar wind is significant for the Sun's long-term evolution. The solar-wind particles begin their journey in the corona in co-rotation with the Sun. At some distance, the particles are released from the strong coronal magnetic fields and then carry a finite azimuthal velocity component into interplanetary space, which is responsible for the particle contribution to the angular-momentum transport \citep{weber67}.  The azimuthal velocity component $U_{\phi}$ of the solar wind decreases with distance from the Sun (assuming a torque-free ballistic trajectory, $U_{\phi}\propto 1/r$), making measurements of $U_{\phi}$ at large heliocentric distances particularly difficult. However, observations of cometary tails suggest non-radial solar-wind velocities \citep{brandt70}, and even early in situ measurements at 1\,au have been used to estimate the Sun's angular-momentum loss \citep{hundhausen70b,lazarus71}.

As the solar wind accelerates from velocities near zero in the Sun's rest frame up to super-sonic, super-Alfv\'enic velocities, it must pass two critical distances: the distance $r_{\mathrm S}$ at which the outflow speed crosses the local sound speed, and  the distance $r_{\mathrm A}$ at which the outflow speed crosses the local Alfv\'en speed \citep{parker58}. Their locations generally depend on heliolatitude and undergo variations depending on the properties of the wind's source regions. The sonic and Alfv\'enic Mach numbers cross the value of unity at these locations, respectively. The heliocentric distance $r_{\beta}$, at which the local thermal pressure in the particles is equal to the energy density in the local magnetic field, is a third important critical distance. All of these critical radii are key predictions of solar-wind models and important for our understanding of the acceleration of the solar wind.

\emph{Parker Solar Probe} and \emph{Solar Orbiter} are the latest additions to the fleet of solar and solar-wind-observing missions \citep{fox16,mueller20}. Both missions carry modern instrumentation  into the inner heliosphere to measure the particles and the electromagnetic fields of the solar wind in situ and monitor the solar-wind outflow remotely. Over the coming years, their observations will improve our understanding of the solar wind at different heliocentric distances and heliolatitudes during solar-minimum and solar-maximum conditions. The goal of our study is the use of radial conservation laws for flux quantities relating to the mass, momentum, energy, and angular momentum of the solar wind to understand their heliolatitudinal variations. Since these quantities are independent of heliocentric distance under a set of assumptions, we use data from \emph{Ulysses} to study their dependence on heliolatitude and solar cycle alone. Within the validity of our assumptions, these measurements serve as contextual information and predictions for the global solar-wind behaviour encountered by \emph{Parker Solar Probe} and \emph{Solar Orbiter}. In addition, we use our scaling parameters to estimate the scaled Alfv\'enic and sonic Mach numbers as well as the critical radii $r_{\mathrm A}$, $r_{\mathrm S}$, and $r_{\beta}$ as functions of heliolatitude based on the \emph{Ulysses} data during solar minimum and solar maximum.  In the future, these scaling laws will be refined with data from \emph{Parker Solar Probe} and \emph{Solar Orbiter}, once both missions have explored a wider range of heliocentric distances and heliolatitudes.

\section{Fluid equations and conservation laws}

 For simplicity, the solar wind is described here as a mostly proton--electron plasma\footnote{This approach neglects the contribution from $\alpha$-particles, which we discuss in Section~\ref{sect:disc}.} with isotropic pressure under the influence of electromagnetic fields. A fluid approach is valid on spatial and temporal scales greater than the characteristic kinetic plasma scales as long as high-order velocity moments of the particle distribution functions can be neglected. We use the proton-fluid continuity equation,
\begin{equation}\label{cont1}
\diffp{N}{t}+\nabla\cdot\left(N\vec U\right)=0,
\end{equation}
and the proton-fluid momentum equation with isotropic, scalar pressure,
\begin{equation}\label{mom1}
NM\left[\diffp{\vec U}{t}+\left(\vec U\cdot \nabla \right) \vec U \right]= -\nabla P  + NQ\left(\vec E+\frac{1}{c}\vec U\times \vec B\right)+NM\vec g,
\end{equation}
where $N$ is the proton density, $\vec U$ is the proton bulk velocity, $M$ is the proton mass, $P$ is the proton pressure, $Q$ is the proton charge, $\vec E$ is the electric field, $c$ is the speed of light, $\vec B$ is the magnetic field, and $\vec g$ is the gravitational acceleration.
The electron-fluid equation, neglecting all terms proportional to the electron mass $m$, is given by
\begin{equation}\label{mom2}
-\nabla p+nq\left(\vec E+\frac{1}{c}\vec u\times \vec B\right)=0,
\end{equation}
where $p$ is the electron pressure, $n$ and $q$ are the electron number density and charge, and $\vec u$ is the electron bulk velocity.
We combine equations~(\ref{mom1}) and (\ref{mom2}) to eliminate $\vec E$. Evoking quasi-neutrality ($N\approx n$), we use the definition of the current density 
\begin{equation}
\vec j=NQ\vec U+nq\vec u
\end{equation}
to obtain
\begin{equation}\label{mom3}
NM\left[\diffp{\vec U}{t}+\left(\vec U\cdot \nabla \right) \vec U\right] = -\nabla \left(P+p\right) + \frac{1}{c}\vec j\times \vec B+NM\vec g.
\end{equation}
Furthermore, we use Amp\`ere's law, $\nabla\times \vec B=4\pi \vec j/c$, to simplify the remaining electromagnetic force terms, and assume steady-state conditions ($\partial/\partial t=0$), leading to
\begin{equation}\label{mom4}
NM\left(\vec U\cdot \nabla \right) \vec U = -\nabla \left(P+p\right)\\
 - \frac{\nabla \vec B^2}{8\pi} +\frac{\left(\vec B\cdot \nabla\right)\vec B}{4\pi}+NM\vec g.
\end{equation}
We now transform equations~(\ref{cont1}) and (\ref{mom4}) into spherical coordinates.  equation~(\ref{cont1}) then yields
\begin{equation}\label{cont2}
\frac{1}{r^2}\diffp{}{r}\left(r^2NU_r\right)+\frac{1}{r\sin\theta}\diffp{}{\theta}\left(NU_{\theta} \sin\theta \right)+\frac{1}{r\sin\theta}\diffp{}{\phi}\left(NU_{\phi}\right)=0,
\end{equation}
where $\theta$ is the polar angle and $\phi$ is the azimuthal angle. In our convention, the heliolatitude $\lambda$ relates to $\theta$ through $\lambda=90^{\circ}-\theta$.
The radial component of equation~(\ref{mom4}) is given by 
\begin{multline}\label{mom2r}
NM \left( U_r\diffp{U_r}{r}+\frac{U_{\theta}}{r}\diffp{U_r}{\theta}+\frac{U_{\phi}}{r\sin\theta}\diffp{U_r}{\phi} - \frac{U_{\theta}^2+U_{\phi}^2}{r} \right)\\
= -\diffp{}{r}\left(P+p+\frac{\vec B^2}{8\pi}\right)\\
+\frac{1}{4\pi}\left(B_r\diffp{B_r}{r}+\frac{B_{\theta}}{r}\diffp{B_r}{\theta}+\frac{B_{\phi}}{r\sin\theta}\diffp{B_r}{\phi}-\frac{B_{\theta}^2+B_{\phi}^2}{r}  \right)-NMg,
\end{multline}
its polar component by
\begin{multline}
NM\left(U_r\diffp{U_{\theta}}{r}+\frac{U_{\theta}}{r}\diffp{U_{\theta}}{\theta}+\frac{U_{\phi}}{r\sin\theta}\diffp{U_{\theta}}{\phi}+\frac{U_rU_{\theta}}{r}-\frac{U_{\phi}^2\mathrm{cot}\theta}{r}\right)\\
=-\frac{1}{r}\diffp{}{\theta}\left(P+p+\frac{\vec B^2}{8\pi}\right)\\
+\frac{1}{4\pi}\left(B_r\diffp{B_{\theta}}{r}+\frac{B_{\theta}}{r}\diffp{B_{\theta}}{\theta}+\frac{B_{\phi}}{r\sin\theta}\diffp{B_{\theta}}{\phi}+\frac{B_rB_{\theta}}{r}-\frac{B_{\phi}^2\mathrm{cot}\theta}{r}\right),
\end{multline}
and its azimuthal component by
\begin{multline}\label{mom2phi}
NM \left(U_r\diffp{U_{\phi}}{r}+\frac{U_{\theta}}{r}\diffp{U_{\phi}}{\theta}+\frac{U_{\phi}}{r\sin\theta}\diffp{U_{\phi}}{\phi}+\frac{U_rU_{\phi}}{r}+\frac{ U_{\phi}U_{\theta}\mathrm{cot}\theta}{r}\right)\\
= -\frac{1}{r\sin\theta}\diffp{}{\phi}\left(P+p+\frac{\vec B^2}{8\pi}\right)\\
+\frac{1}{4\pi}\left(B_r\diffp{B_{\phi}}{r}+\frac{B_{\theta}}{r}\diffp{B_{\phi}}{\theta}+\frac{B_{\phi}}{r\sin\theta}\diffp{B_{\phi}}{\phi}+\frac{B_rB_{\phi}}{r}+\frac{B_{\phi}B_{\theta}\mathrm{cot}\theta}{r}\right).
\end{multline}
We now assume azimuthal symmetry ($\partial/\partial \phi=0$). Although the observed solar wind exhibits a non-zero polar component $U_{\theta}$ of the bulk velocity and a non-zero polar component $B_{\theta}$ of the magnetic field at times, we assume that $U_{\theta}=B_{\theta}=0$ on average as in the \citet{parker58} model.  The condition $\nabla\cdot \vec B=0$ under our assumptions reduces to
\begin{equation}\label{divB}
\diffp{}{r}\left(r^2B_r\right)=0.
\end{equation}
Likewise, continuity according to equation~(\ref{cont2}) simplifies to
\begin{equation}\label{continuity}
\diffp{\mathcal F_m}{r}=0,
\end{equation}
where
\begin{equation}
\mathcal F_{m}=r^2NMU_r
\end{equation}
is the radial mass flux per steradian. The momentum equations in equations~(\ref{mom2r}) through (\ref{mom2phi}) simplify to
\begin{equation}\label{mom3r}
NM \left( U_r\diffp{U_r}{r}- \frac{U_{\phi}^2}{r} \right)= -\diffp{}{r}\left(P+p\right)-\frac{1}{r^2}\diffp{}{r}\left(r^2\frac{B_{\phi}^2}{8\pi}\right)-NMg,
\end{equation}
\begin{equation}\label{momtheta}
NMU_{\phi}^2=\frac{1}{4\pi}B_{\phi}^2,
\end{equation}
and
\begin{equation}\label{mom3phi}
NM \left(U_r\diffp{U_{\phi}}{r}+\frac{U_rU_{\phi}}{r}\right)= \frac{1}{4\pi}\left(B_r\diffp{B_{\phi}}{r}+\frac{B_rB_{\phi}}{r}\right).
\end{equation}  

By combining equation~(\ref{continuity}) with equations~(\ref{mom3r}) and (\ref{momtheta}), we find momentum conservation in the form
\begin{equation}\label{diffFp}
\diffp{\mathcal F_{p}}{r}=-r^2 \diffp{}{r}\left(P+p+\frac{B_{\phi}^2}{8\pi}\right)-NMGM_{\odot},
\end{equation}
where
\begin{equation}
\mathcal F_{p}=r^2NMU_r^2
\end{equation}
is the radial kinetic momentum flux per steradian,  $G$ is the gravitational constant, and $M_{\odot}$ is the Sun's mass. The right-hand side of equation~(\ref{diffFp}) is zero if the solar wind is ``coasting'' without radial acceleration, which is a reasonable assumption for heliocentric distances greater than about 0.3 au, especially in fast wind \citep{marsch84}. Slow wind, however, still experiences some acceleration to distances $\gtrsim 1\,\mathrm{au}$ \citep{schwenn81}. Although this acceleration effect is small, we urge caution when extending our framework into the inner heliosphere. Even if the coasting approximation (i.e., $\mathcal F_{\mathrm p}=\text{constant}$) is not fulfilled at the location of the measurement, $\mathcal F_p$ describes the radial component of the particle momentum at this location; however, the scaling of this quantity to different radial distances requires the inclusion of the right-hand side of equation~(\ref{diffFp}).

Under the same assumptions, we find that
\begin{multline}\label{diffFe}
\diffp{}{r}\left[r^2NMU_r\left(\frac{\vec U^2}{2}+\frac{\gamma}{\gamma-1}\frac{P+p}{NM}-\frac{GM_{\odot}}{r} \right)\right]\\
=\frac{r}{4\pi}\left(U_{\phi}B_r-U_rB_{\phi}\right)\diffp{}{r}\left(rB_{\phi}\right),
\end{multline}
where $\gamma$ is the polytropic index (assumed to be equal for protons and electrons).  In the derivation of equation~(\ref{diffFe}), we use the polytopic assumption for both protons and electrons:
\begin{equation}
P\propto N^{\gamma}
\end{equation}
and
\begin{equation}
p\propto n^{\gamma}. 
\end{equation}
We note that equation~(\ref{diffFe}) can be easily extended to account for different polytropic indices for protons and electrons.
In order to simplify the right-hand side of equation~(\ref{diffFe}), we evoke the frozen-in condition of the magnetic field \citep{parker58}. 
In a frame that co-rotates with the Sun, the magnetic field lines are parallel to $\vec U$  \citep{weber67,mestel68,verscharen15}. This condition leads to 
\begin{equation}\label{BphiBr}
\frac{B_{\phi}}{B_r}=\frac{U_{\phi}-\Omega_{\odot}r\sin \theta}{U_r},
\end{equation}
where $\Omega_{\odot}$ is the Sun's angular rotation frequency, which we assume to be constant for all $\theta$. From equation~(\ref{BphiBr}), we find the useful identity
\begin{equation}
r\left(U_{\phi}B_r-U_rB_{\phi}\right) =r^2 B_r\Omega_{\odot}\sin \theta=\mathrm{constant},
\end{equation}
where the second equality follows from equation~(\ref{divB}). This relationship allows us to simplify the right-hand side of equation~(\ref{diffFe}) so that the energy-conservation law yields
\begin{equation}
\diffp{\mathcal F_E}{r}=0,
\end{equation}
where
\begin{multline}\label{Feeq}
\mathcal F_{E}=r^2NMU_r\left(\frac{\vec U^2}{2} +\frac{\gamma}{\gamma-1}\frac{P+p}{NM} -\frac{GM_{\odot}}{r} \right.\\
\left. - \frac{r B_rB_{\phi}}{4\pi NM U_r}\Omega_{\odot}\sin \theta \right)
\end{multline}
is the radial energy flux per steradian.

By combining equations~(\ref{divB}) and (\ref{continuity}) with equation~(\ref{mom3phi}), we furthermore identify angular-momentum conservation in the form
\begin{equation}
\diffp{\mathcal F_L}{r}=0,
\end{equation}
where
\begin{equation}
\mathcal F_{L}=r^3NMU_rU_{\phi}-r^3\frac{B_rB_{\phi}}{4\pi}.
\end{equation}

The quantities $\mathcal F_m$, $\mathcal F_p$ (within the coasting approximation), $\mathcal F_E$, and $\mathcal F_L$ are constant with heliocentric distance. We note that, albeit useful for the description of averaged and global-scale variations, this model ignores any variations due to asymmetries, stream interactions, and natural fluctuations (for a further discussion of these effects, see Section~\ref{sect:disc}).

\section{Data analysis}

We use 30-h averages of the proton and magnetic-field data recorded by \textit{Ulysses} during its three polar orbits.  We choose an average interval of 30~h to sample over time-scales that are greater than the typical correlation time of the ubiquitous solar-wind fluctuations \citep[typically of order a few hours;][]{matthaeus82,bruno86a,tu95,damicis10,bruno13}. At the same time, this averaging interval is short enough to avoid significant  variations in \emph{Ulysses'} heliolatitude during the  recording of each data point in our averaged data set.  The proton measurements were recorded by the Solar Wind Observations Over the Poles of the Sun (SWOOPS) instrument \citep{bame92}. The magnetic-field measurements were recorded by the  Magnetic Field experiment \citep{balogh92}. In order to visualise the differences between solar minimum and solar maximum conditions, we only use data from \emph{Ulysses'} three fast heliolatitude scans. The first scan occurred during solar minimum, the second scan occurred during solar maximum, and the third scan occurred during the following (deep) solar minimum \citep{mccomas08}. We select data from DOY 256 in 1994 until DOY 213 in 1995 (known as Fast Latitude Scan 1, FLS1) and data from DOY 38 in 2007 until DOY 13 in 2008 (known as Fast Latitude Scan 3, FLS3), and label these data as ``solar minimum''. We select data from DOY 329 in 2000 until DOY 285 in 2001 (known as Fast Latitude Scan 2, FLS2) and label these data as ``solar maximum''.  During these time intervals, \emph{Ulysses'} eccentric orbit brought the spacecraft to heliocentric distances between 1.34\,au at the perihelia and 2.37\,au at the furthest polar pass.
We summarise our results in Table~\ref{tab_summary}.

\begin{table*}
 \contcaption{Summary of our measurement results. We show the mean values, minimum values, and maximum values of the quantities illustrated in Figures~\ref{fig:dial_massflux} through \ref{fig:dial_rBeta} for the three fast latitudinal scans (FLSs). The given error bars of the mean values represent the calculated standard errors of the mean.}
 \label{tab_summary}
 \begin{tabular}{lccccccccc}
  \hline
  Quantity & \multicolumn{3}{c}{FLS1 (solar minimum)} & \multicolumn{3}{c}{FLS2 (solar maximum)} & \multicolumn{3}{c}{FLS3 (solar minimum)}\\
   & Mean & Min & Max & Mean & Min & Max & Mean & Min & Max\\
  \hline
$\mathcal F_m\,(10^{-16}\,\mathrm{au}^2\,\mathrm{g\,cm}^{-2}\,\mathrm s^{-1}\,\mathrm{sr}^{-1})$ & 3.549$\pm$0.069 & 1.75 & 11.06 &  4.486$\pm$0.169 & 0.680 & 21.30 &  2.682 $\pm$0.082  & 0.958 & 15.44  \\
$\mathcal F_p\,(10^{-8}\,\mathrm{au}^2\,\mathrm {g\,cm}^{-1}\,\mathrm s^{-2}\,\mathrm{sr}^{-1})$ & 2.426$\pm$0.029 & 0.739 & 4.04 & 2.075$\pm$0.076 & 0.349 & 9.81 & 1.705$\pm$0.038 & 0.542 & 7.97 \\
$\mathcal F_E\,(\mathrm{au}^2\,\mathrm g\,\mathrm s^{-3}\,\mathrm{sr}^{-1})$ & 0.913 $\pm$ 0.015 & 0.166 & 1.35 & 0.554$\pm$0.025 & 0.076 & 2.64  & 0.601$\pm$0.014 & 0.097 & 2.25 \\
$\mathcal F_L\,(10^{-9}\,\mathrm{au}^3\,\mathrm {g\,cm}^{-1}\,\mathrm s^{-2}\,\mathrm{sr}^{-1})^{\ast}$ & 1.162$\pm$0.037  & 0.060 & 1.94 & 0.715$\pm$0.074 & 0.001 & 3.58 & 0.765$\pm$ 0.026 & 0.018 & 1.38 \\
$\tilde M_{\mathrm A}$ & 18.892$\pm$1.426 & 7.23 & 343.43 & 21.339$\pm$1.597 & 4.21 & 200.44 & 22.300$\pm$ 0.541 & 6.54 & 99.00 \\
$r_{\mathrm A}\,(R_{\odot})$ & 12.080$\pm$0.236 & 0.128 & 26.00 & 12.766$\pm$ 0.522& 0.008 & 47.15 & 9.504$\pm$0.221 & 0.289 & 28.31 \\
$\tilde M_{\mathrm S}$ & 11.409$\pm$0.057 & 7.87 & 16.93 & 11.594$\pm$0.148 & 7.04 & 18.72 & 12.079$\pm$0.079 & 7.69 & 17.96 \\
$r_{\mathrm S}\,(R_{\odot})$ & 0.309$\pm$0.004 & 0.104 & 0.819 & 0.346$\pm$0.012 & 0.079 & 1.11 & 0.273$\pm$0.006 & 0.088 & 0.871 \\
$r_{\beta}\,(\mathrm{au})^{\dagger}$ & 0.552$\pm$0.028 & 0.011 & 3.18 & 0.658$\pm$0.052 & 0.000 & 4.07 & 0.488$\pm$0.048 & 0.036 & 6.69 \\
  \hline
 \end{tabular}
 
\begin{flushleft}
$^\ast$ The statistics for $\mathcal F_L$ only include those times when $\mathcal F_L>0$.

$^\dagger$ The statistics for $r_{\beta}$ only include those times when $r_{\beta}>0$.
\end{flushleft}
\end{table*}

\subsection{Mass, momentum, energy, and angular-momentum flux}

\begin{figure}
	\includegraphics[width=\columnwidth]{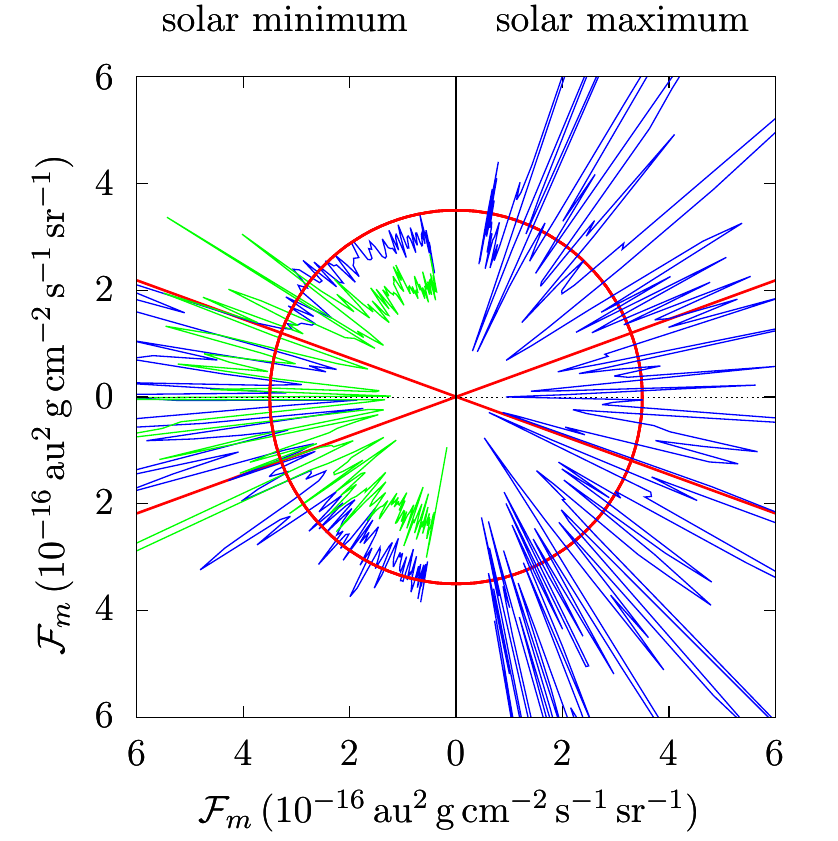}
    \caption{Polar plot of the radial mass flux per steradian $\mathcal F_m$. The polar angle represents the heliolatitude $\lambda$ at which \textit{Ulysses} recorded the measurement. The distance from the centre of the plot describes the local value of $\mathcal F_m$. The red lines indicate $\lambda=\pm 20^{\circ}$. The red circle has a radius of $3.5\times10^{-16}\,\mathrm{au}^2\,\mathrm{g\,cm}^{-2}\,\mathrm s^{-1}\,\mathrm{sr}^{-1}$. The left half of the figure shows conditions during solar minimum from FLS1 (blue) and FLS3 (green), and the right half of the figure shows conditions during solar maximum from FLS2 (blue).}
  \label{fig:dial_massflux}
\end{figure}

 In Figure~\ref{fig:dial_massflux}, we show a polar plot of  the radial mass flux per steradian $\mathcal F_m$ based on the \emph{Ulysses} measurements. The polar angle in this diagram and in the following diagrams illustrates the heliolatitude at which the measurement was taken. The red lines indicate the heliolatitudes of $\pm 20^{\circ}$, which \citet{mccomas00} identify as the separation between slow equatorial streamer-belt wind and fast polar coronal-hole wind during solar minimum. The red circle indicates a constant value of $\mathcal F_m=3.5\times10^{-16}\,\mathrm{au}^2\,\mathrm{g\,cm}^{-2}\,\mathrm s^{-1}\,\mathrm{sr}^{-1}$ and is meant as a help to guide the eye.
During solar minimum, $\mathcal F_m$ varies between about $1\times 10^{-16}$ and $15\times 10^{-16}\,\mathrm{au}^2\,\mathrm{g\,cm}^{-2}\,\mathrm s^{-1}\,\mathrm{sr}^{-1}$ in the equatorial region, while it is steadier over the polar regions beyond $\pm 20^{\circ}$ at a value of about $3.5\times 10^{-16}\,\mathrm{au}^2\,\mathrm{g\,cm}^{-2}\,\mathrm s^{-1}\,\mathrm{sr}^{-1}$ during FLS1. The polar mass flux is lower at a value of about $2.2\times 10^{-16}\,\mathrm{au}^2\,\mathrm{g\,cm}^{-2}\,\mathrm s^{-1}\,\mathrm{sr}^{-1}$ during FLS3.  During solar maximum, $\mathcal F_m$ exhibits large variations consistent with the larger variability of the solar-wind source regions. The maximum value during solar maximum is about $2.1\times 10^{-15}\,\mathrm{au}^2\,\mathrm{g\,cm}^{-2}\,\mathrm s^{-1}\,\mathrm{sr}^{-1}$, which is almost by a factor of 10 greater than the average value over polar regions during solar minimum. The clear separation between equatorial and polar wind vanishes during solar maximum.

\begin{figure}
	\includegraphics[width=\columnwidth]{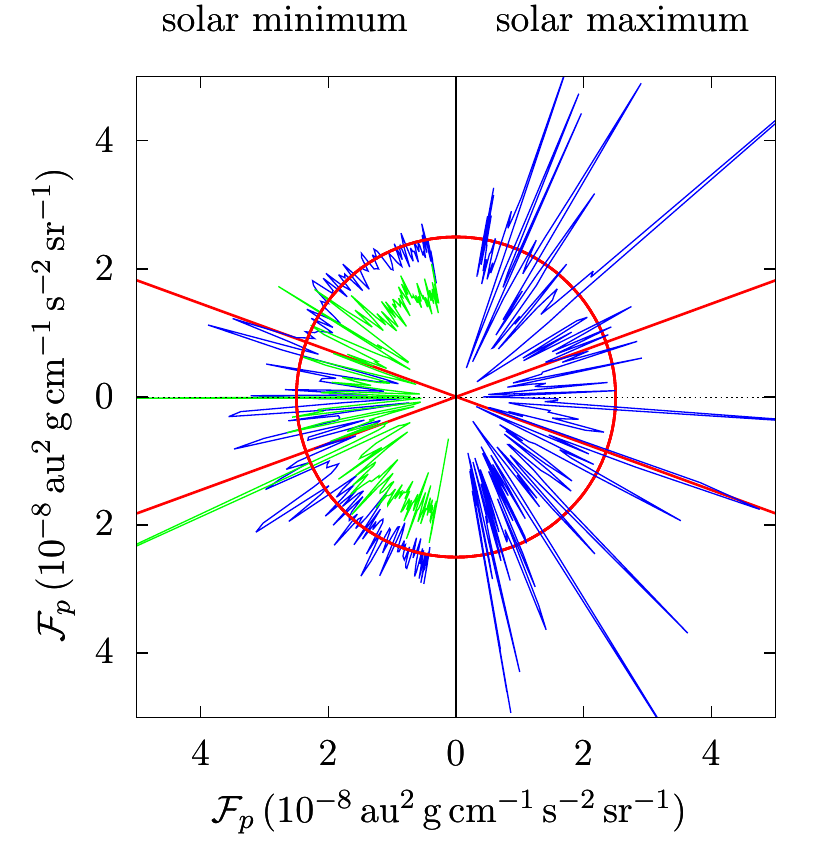}
    \caption{Polar plot of the radial particle momentum flux per steradian $\mathcal F_p$ during solar minimum (left half, FLS1 in blue, FLS3 in green) and solar maximum (right half, FLS2). The format of this plot is the same as in Figure~\ref{fig:dial_massflux}. The red circle has a radius of $2.5\times10^{-8}\,\mathrm{au}^2\,\mathrm{g\,cm}^{-1}\,\mathrm s^{-2}\,\mathrm{sr}^{-1}$.}
  \label{fig:dial_momflux}
\end{figure}

We show the polar plot of the radial particle momentum flux per steradian $\mathcal F_p$ in Figure~\ref{fig:dial_momflux} for solar-minimum and solar-maximum conditions. During solar minimum, $\mathcal F_p$ presents variations between about $0.5\times 10^{-8}$ and $8\times 10^{-8}\,\mathrm{au}^2\,\mathrm{g\,cm}^{-1}\,\mathrm s^{-2}\,\mathrm{sr}^{-1}$ at equatorial heliolatitudes below $\pm 20^{\circ}$. Outside the equatorial region, $\mathcal F_p$ is almost independent of heliolatitude at a value of approximately $2.5\times 10^{-8}\,\mathrm{au}^2\,\mathrm{g\,cm}^{-1}\,\mathrm s^{-2}\,\mathrm{sr}^{-1}$ during FLS1 and  at a value of approximately $1.7\times 10^{-8}\,\mathrm{au}^2\,\mathrm{g\,cm}^{-1}\,\mathrm s^{-2}\,\mathrm{sr}^{-1}$ during FLS3. Like in the case of $\mathcal F_m$, also $\mathcal F_p$ shows a strong variation during solar maximum between values from less than $3\times 10^{-9}$ to almost $10^{-7}\,\mathrm{au}^2\,\mathrm{g\,cm}^{-1}\,\mathrm s^{-2}\,\mathrm{sr}^{-1}$ at times. At equatorial heliolatitudes, the average $\mathcal F_p$ does not differ much between solar minimum and solar maximum, although its variability is greater during solar maximum. 

\begin{figure}
	\includegraphics[width=\columnwidth]{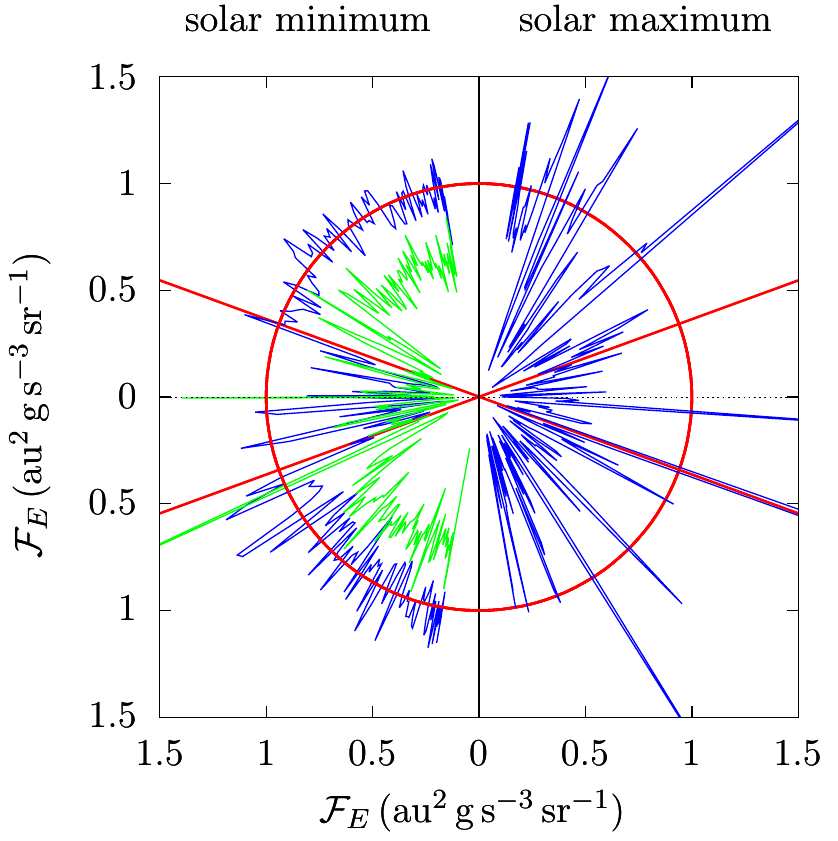}
    \caption{Polar plot of the radial energy flux per steradian $\mathcal F_E$ during solar minimum (left half, FLS1 in blue, FLS3 in green) and solar maximum (right half, FLS2). We use $\gamma=5/3$ and assume that $p=P$. The format of this plot is the same as in Figure~\ref{fig:dial_massflux}. The red circle has a radius of $1\,\mathrm{au}^2\,\mathrm{g\,s}^{-3}\,\mathrm{sr}^{-1}$.}
  \label{fig:dial_energflux}
\end{figure}

Figure~\ref{fig:dial_energflux} shows our polar plot of the radial energy flux per steradian $\mathcal F_E$. During solar minimum, $\mathcal F_E$ exhibits a significant difference between equatorial and polar regions. Near the equator between $\lambda=\pm 20^{\circ}$, we observe $\mathcal F_E$ between about 0.1 and $2.2\,\mathrm{au}^2\,\mathrm{g}\,\mathrm s^{-3}\,\mathrm{sr}^{-1}$. Outside the equatorial heliolatitudes, we observe an average value of about $1\,\mathrm{au}^2\,\mathrm{g}\,\mathrm s^{-3}\,\mathrm{sr}^{-1}$ during FLS1 and about $0.7\,\mathrm{au}^2\,\mathrm{g}\,\mathrm s^{-3}\,\mathrm{sr}^{-1}$ during FLS3, independent of $\lambda$. During solar maximum, $\mathcal F_E$ expectedly shows a larger variability between values from below 0.1 to above $2.6\,\mathrm{au}^2\,\mathrm{g}\,\mathrm s^{-3}\,\mathrm{sr}^{-1}$. At the location of the measurement, $\mathcal F_E$ is dominated by the kinetic-energy contribution of the protons.

\begin{figure}
	\includegraphics[width=\columnwidth]{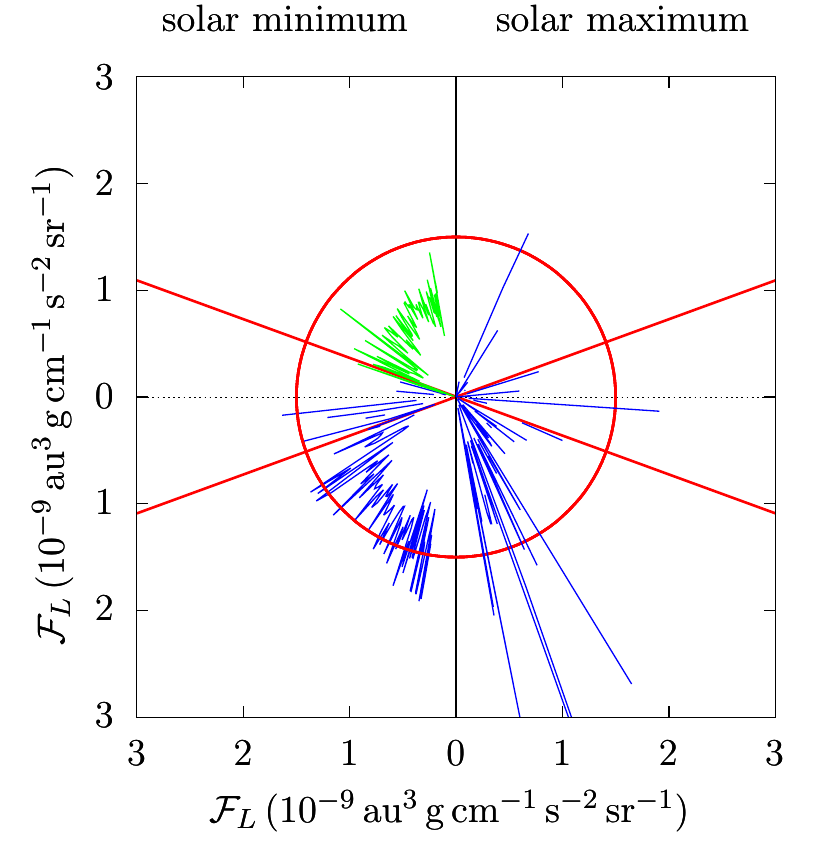}
    \caption{Polar plot of the radial angular-momentum flux per steradian $\mathcal F_L$ during solar minimum (left half, FLS1 in blue, FLS3 in green) and solar maximum (right half, FLS2). We only plot $\mathcal F_L$ if $\mathcal F_{L}>0$. The format of this plot is the same as in Figure~\ref{fig:dial_massflux}. The red circle has a radius of $1.5\times 10^{-9}\,\mathrm{au}^3\,\mathrm{g\,cm}^{-1}\,\mathrm s^{-2}\,\mathrm{sr}^{-1}$.}
  \label{fig:dial_angmomflux}
\end{figure}

We show the polar plot of the radial angular-momentum flux per steradian $\mathcal F_L$ in Figure~\ref{fig:dial_angmomflux}. Due to pointing uncertainties in the \emph{Ulysses} data set (for details, see Section~\ref{sect:disc}), the measurement of $U_{\phi}$ is prone to a much larger uncertainty than the measurement of $U_r$. We, therefore, only plot $\mathcal F_L$ when $\mathcal F_L>0$. Due to the data gaps when $\mathcal F_L<0$, it is impossible to define a meaningful average value for $\mathcal F_L>0$ at Northern heliolatitudes above the equatorial plane during FLS1 and at Southern heliolatitudes below the equatorial plane during FLS3. Even the equatorial values need to be treated with caution. During solar minimum, we find that $\mathcal F_L$ is approximately $1.2\times 10^{-9}\,\mathrm{au}^3\,\mathrm{g\,cm}^{-1}\,\mathrm s^{-2}\,\mathrm{sr}^{-1}$ in the Southern polar region (FLS1) and approximately $0.8\times 10^{-9}\,\mathrm{au}^3\,\mathrm{g\,cm}^{-1}\,\mathrm s^{-2}\,\mathrm{sr}^{-1}$ in the Northern polar region (FLS3). During solar maximum, its value varies between $6\times 10^{-12}\,\mathrm{au}^3\,\mathrm{g\,cm}^{-1}\,\mathrm s^{-2}\,\mathrm{sr}^{-1}$ and about $3.6\times 10^{-9}\,\mathrm{au}^3\,\mathrm{g\,cm}^{-1}\,\mathrm s^{-2}\,\mathrm{sr}^{-1}$. However, we re-iterate that these values need to be treated with caution based on the pointing uncertainty of the spacecraft.

\subsection{Alfv\'enic Mach number and the Alfv\'en radius}\label{scalings}

In this section, we derive the value of the Alfv\'enic Mach number scaled to a heliocentric distance of 1\,au and the location of the Alfv\'en radius as functions of heliolatitude. For this calculation, we require a scaling law for the magnetic field $\vec B$. Throughout this work, we use the tilde symbol to indicate a quantity that has been scaled to its value at a heliocentric distance of 1\,au.
Using assumptions consistent with ours, \citet{parker58} provides expressions for the averaged global-scale heliospheric magnetic field as
\begin{equation}\label{Parker1}
B_r(r)=B_r(r_0)\left(\frac{r_0}{r}\right)^2,
\end{equation}
\begin{equation}
B_{\theta}=0,
\end{equation}
and
\begin{equation}\label{Parker3}
B_{\phi}(r)=B_r(r)\frac{\Omega_{\odot}\sin\theta }{U_r}\left(b-r\right),
\end{equation}
where $r_0$ is any arbitrary reference distance from the Sun and $b$ is the effective source-surface radius. 
In order to scale the measured magnetic field from the \emph{Ulysses} data set to its value at 1\,au, we assume that the averaged heliospheric magnetic field follows, to first order, the Parker magnetic field.  Since $b\ll r$, we approximate equation~(\ref{Parker3}) as
\begin{equation}\label{Bphiapp}
B_{\phi}(r)\approx B_{\phi}(r_0)\left(\frac{r_0}{r}\right).
\end{equation}
Using equations~(\ref{Parker1}) and (\ref{Bphiapp}), we approximate the magnitude of the scaled magnetic field at a heliocentric distance of 1\,au as
\begin{equation}\label{Bscaled}
\tilde B=\left(\frac{r}{1\,\mathrm{au}}\right) \sqrt{\left(\frac{r}{1\,\mathrm{au}}\right)^2B_r^2+B_{\phi}^2},
\end{equation}
where $B_r$ and $B_{\phi}$ are the measured magnetic-field components at the heliocentric distance $r$, and $r$ is the heliocentric distance of the measurement location. 
Assuming that the proton bulk velocity remains independent of $r$ for $r\gtrsim 1\,\mathrm{au}$, equation~(\ref{continuity}) suggests that $N\propto r^{-2}$, allowing us to define the scaled proton density at a heliocentric distance of 1\,au as
\begin{equation}\label{nscaled}
\tilde N= N\left(\frac{r}{1\,\mathrm{au}}\right)^2.
\end{equation}
Using equations~(\ref{Bscaled}) and (\ref{nscaled}), we define the scaled Alfv\'en speed at a heliocentric distance of 1\,au as
\begin{equation}\label{vAscaled}
\tilde v_{\mathrm A}=\frac{\tilde B}{\sqrt{4\pi \tilde NM}}.
\end{equation}
Likewise, we define the scaled Alfv\'enic Mach number at a heliocentric distance of 1\,au as
\begin{equation}
\tilde M_{\mathrm A}=\frac{U_r}{\tilde v_{\mathrm A}},
\end{equation}
again relying on the assumption that $U_r$ is independent of $r$ for $r\gtrsim 1\,\mathrm{au}$. 

\begin{figure}
	\includegraphics[width=\columnwidth]{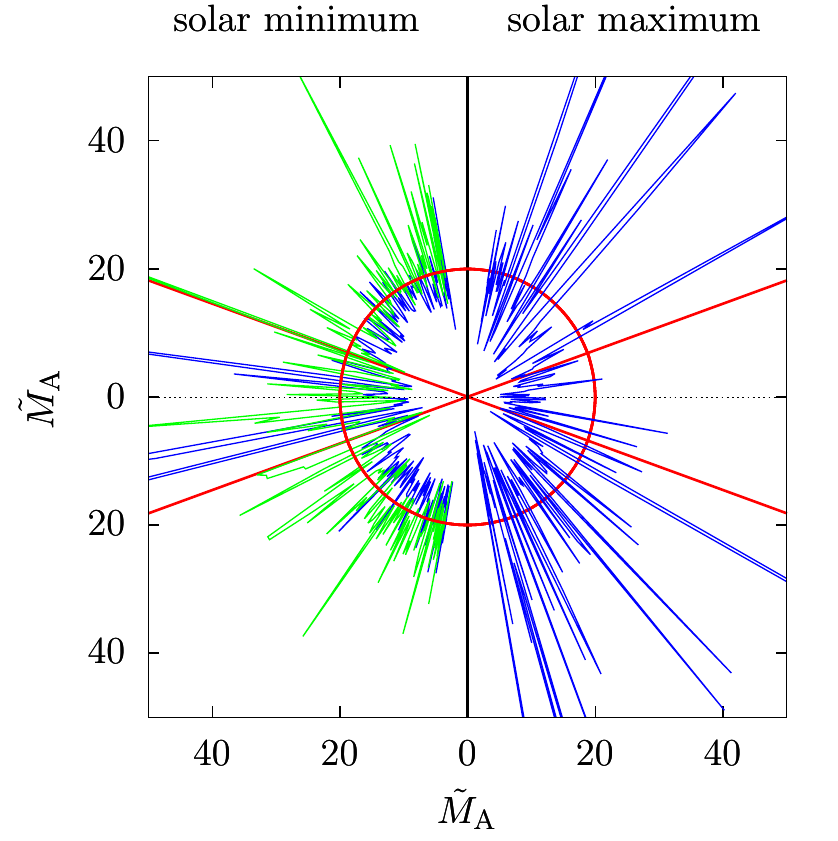}
    \caption{Polar plot of the scaled Alfv\'enic Mach number $\tilde M_{\mathrm A}$ at $r=1\,\mathrm{au}$ during solar minimum (left half, FLS1 in blue, FLS3 in green) and solar maximum (right half, FLS2). The format of this plot is the same as in Figure~\ref{fig:dial_massflux}. The red circle has a radius of 20.}
  \label{fig:dial_Ma}
\end{figure}

We show our polar plot of the scaled Alfv\'enic Mach number $\tilde M_{\mathrm A}$ in Figure~\ref{fig:dial_Ma}. The scaled solar wind at 1\,au is super-sonic ($\tilde M_{\mathrm A}>1$) at all heliolatitudes and both during solar minimum and solar maximum. During solar minimum, $\tilde M_{\mathrm A}$ exhibits more variation at equatorial heliolatitudes with values between 7 and 343. Outside the equatorial region, the solar-minimum value of $\tilde M_{\mathrm A}$ varies between 12 and 30 during FLS1 and between 13 and 57 during FLS3. On average, $\tilde M_{\mathrm A}$ is greater during FLS3 than during FLS1. We observe a slight increase of $\tilde M_{\mathrm A}$ with increasing $|\lambda|$ even above $\pm 20^{\circ}$. During solar maximum, $\tilde M_{\mathrm A}$ exhibits a large variability between values from about 4 to extreme cases with values over 200 at times. During solar maximum, the maxima and the variations in $\tilde M_{\mathrm A}$ are greater in polar regions than near the equator.

\begin{figure}
	\includegraphics[width=\columnwidth]{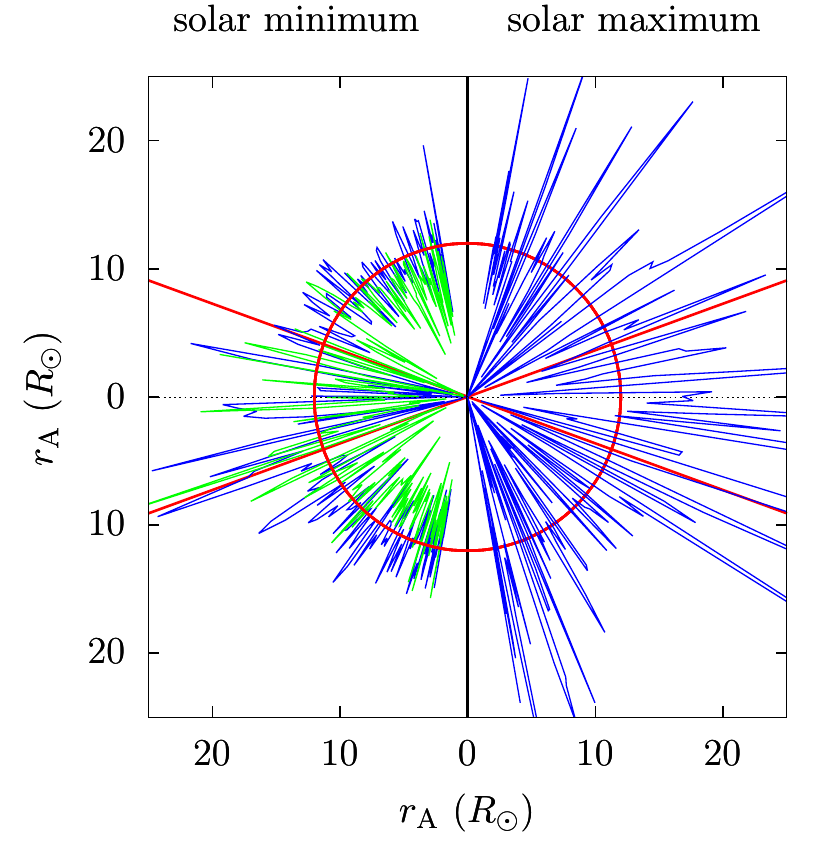}
    \caption{Polar plot of the estimated Alfv\'en radius $r_{\mathrm A}$ based on the scaled magnetic field and proton density during solar minimum (left half, FLS1 in blue, FLS3 in green) and solar maximum (right half, FLS2). The values of $r_{\mathrm A}$ give lower estimates for the Alfv\'en radius. We only plot $r_{\mathrm A}$ if $r_{\mathrm A}>0$. The format of this plot is the same as in Figure~\ref{fig:dial_massflux}. The red circle has a radius of $12R_{\odot}$.}
  \label{fig:dial_rA}
\end{figure}

The Alfv\'en radius $r_{\mathrm A}$ is generally defined as the heliocentric distance $r$, at which the radial proton bulk velocity fulfils 
\begin{equation}\label{Alfvpoint}
U_r=v_{\mathrm A}(r),
\end{equation}
where $v_{\mathrm A}(r)$ is the local Alfv\'en speed. Our scaling assumptions require that $b\ll r$ and that $U_r$ is constant with distance from the Sun as in the original \citet{parker58} model for the interplanetary magnetic field. We now extend the assumption that $\partial U_r/\partial r=0$ to all distances $r\gtrsim r_{\mathrm A}$. We recognize that this assumption is sometimes violated. It allows us, however, to set a reasonable upper limit on $U_r$ as a function of $r$. As long as the actual $U_r(r_{\mathrm A})$ is less than the measured $U_r$ at distance $r$ and $v_{\mathrm A}$ is a monotonic function of $r$, the extension of our scaling relations in equations~(\ref{Bscaled}) and (\ref{nscaled}) to $r=r_{\mathrm A}$ provides us then with a lower-limit estimate for the Alfv\'en radius. Using the condition in equation~(\ref{Alfvpoint}), we find 
\begin{equation}\label{rA}
r_{\mathrm A}=r\sqrt{\frac{B_r^2}{4\pi NM U_r^2-B_{\phi}^2}},
\end{equation}
where $B_r$, $B_{\phi}$, $N$, and $U_r$ are the measured quantities at heliocentric distance $r$.  We show the polar plot of the estimated Alfv\'en radius $r_{\mathrm A}$ according to equation~(\ref{rA}) in Figure~\ref{fig:dial_rA}. During solar minimum, $r_{\mathrm A}$ exhibits more variation in the equatorial region compared to the polar region. Within heliolatitudes of $\pm 20^{\circ}$ of the equator, $r_{\mathrm A}$ varies between about $0.1R_{\odot}$ and $28R_{\odot}$. Polewards from this equatorial region, we find $r_{\mathrm A}$ between about $7R_{\odot}$ and $16R_{\odot}$ with a mean of approximately $12R_{\odot}$ during FLS1. During FLS3,   $r_{\mathrm A}$ is on average smaller in the polar regions with a mean value of approximately $10R_{\odot}$. As expected, the value of $r_{\mathrm A}$ exhibits more variability during solar maximum between values from less than $0.01R_{\odot}$ to $47R_{\odot}$ in extreme cases. We note that values of $r_{\mathrm A}<1R_{\odot}$ are unphysical as these would lie within the sphere of the Sun. 

\subsection{Sonic Mach number, the sonic radius, and the $\beta=1$ radius}\label{sect:sonic}

In this section, we derive the value of the sonic Mach number scaled to a heliocentric distance of 1\,au, the location of the sonic radius, and the location of the $\beta=1$ radius as functions of heliolatitude. For this calculation, we require a scaling law for the proton temperature $T$. 
Fits to the proton temperature profiles observed by \emph{Ulysses} during its first polar orbit reveal a temperature dependence on heliocentric distance and heliolatitude of the form \citep{mccomas00}
\begin{equation}\label{Tpscale}
T=\left[2.58\times 10^5\,\mathrm K+(223\,\mathrm K)\left(\frac{\lambda}{1^{\circ}}\right) \right]\left(\frac{r}{1\,\mathrm{au}}\right)^{-1.02}
\end{equation}
at high heliolatitudes ($|\lambda|\geq 36^{\circ}$). The temperature distribution at equatorial heliolatitudes during solar minimum and, in general, during solar maximum is more complex \citep{mccomas02}. Nevertheless, we apply the power-law scaling in equation~(\ref{Tpscale}) to approximately scale $T$ at the location of measurement back to its value at 1\,au as
\begin{equation}\label{Tp}
\tilde T=T\left(\frac{r}{1\,\mathrm{au}}\right)^{1.02}.
\end{equation}
The \emph{Ulysses} data set provides us with proton temperature measurements achieved in two different ways: one data product corresponds to the integrated second velocity moment of the three-dimensional velocity distribution function. The other data product corresponds to the sum of the second-order moments of one-dimensional energy spectra, avoiding any channels above the proton peak to avoid contamination by $\alpha$-particles. These one-dimensional spectra are calculated as the sum of the measurements over all angles at each fixed energy. Unless the solar-wind temperature $T$ assumes extreme values, the integrated second velocity moment of the three-dimensional velocity distribution function is expected to provide an upper limit on $T$. In these cases, the sum of the second-order moments of the one-dimensional energy spectra is expected to provide a lower limit on $T$.  We use the arithmetic mean of the time averages of both data products as our value of $T$.

\begin{figure}
\includegraphics[width=\columnwidth]{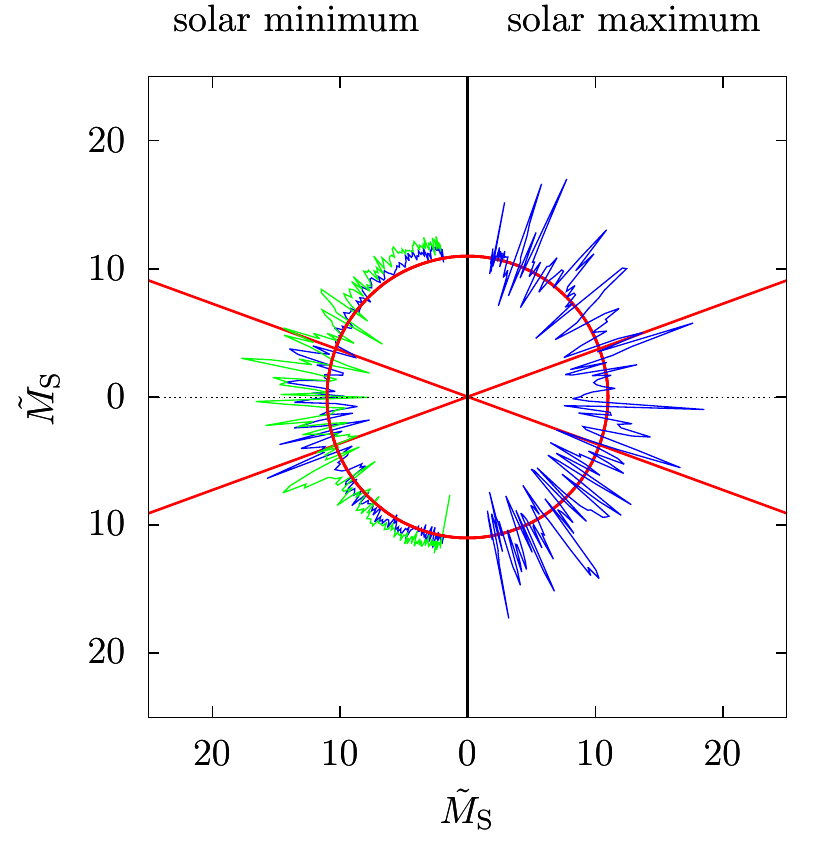}
    \caption{Polar plot of the scaled sonic Mach number $\tilde M_{\mathrm S}$ at $r=1\,\mathrm{au}$ during solar minimum (left half, FLS1 in blue, FLS3 in green) and solar maximum (right half, FLS2). The format of this plot is the same as in Figure~\ref{fig:dial_massflux}. The red circle has a radius of 11.}
  \label{fig:dial_Ms}
\end{figure}

We define the scaled sound speed at a heliocentric distance of 1\,au as
\begin{equation}\label{cS}
\tilde c_{\mathrm S}=\sqrt{\frac{\gamma k_{\mathrm B}\tilde T}{M}},
\end{equation}
where $k_{\mathrm B}$ is the Boltzmann constant. For the sake of simplicity, we set $\gamma=5/3$. This definition allows  us to introduce the scaled sonic Mach number at a heliocentric distance of 1\,au as
\begin{equation}
\tilde M_{\mathrm S}=\frac{U_r}{\tilde c_{\mathrm S}}
\end{equation}
under the assumption that $\partial U_r/\partial r=0$ at distances $r\gtrsim 1\,\mathrm{au}$.
Figure~\ref{fig:dial_Ms} displays the polar plot of the scaled sonic Mach number $\tilde M_{\mathrm S}$. The value of $\tilde M_{\mathrm S}$ shows the least relative variability compared to the other quantities shown in this work throughout the three polar passes that we study. During solar minimum, $\tilde M_{\mathrm S}$ varies between 8 and 18. At $|\lambda|>20^{\circ}$, $\tilde M_{\mathrm S}$ is approximately constant at a value of 11 during FLS1 and at a value of 12 during FLS3. During solar maximum, $\tilde M_{\mathrm S}$ exhibits moderate variations with values between approximately 7 and 19. The difference between equatorial and polar wind is less pronounced during solar maximum.

Assuming an average radial scaling of $T$ at heliocentric distances $<1\,\mathrm{au}$ in addition to the scaling in equation~(\ref{Tp}), which is valid at heliocentric distances $>1\,\mathrm{au}$, allows us to estimate the value of the sonic radius $r_{\mathrm S}$ based on the \emph{Ulysses} measurements depending on heliographic latitude in cases when $r_{\mathrm S}<1\,\mathrm{au}$. The sonic radius is defined as the heliocentric distance $r$, at which the radial proton bulk velocity fulfils
\begin{equation}\label{soniccond}
U_r=c_{\mathrm S}(r),
\end{equation}
where $c_{\mathrm S}(r)$ is the local sound speed. Like in the case of the Alfv\'en radius, we apply our assumption that $\partial U_r/\partial r=0$ to all distances $r\gtrsim r_{\mathrm S}$, an assumption that is prone to the same caveats as in the case of $r_{\mathrm A}$. Fits to radial profiles of $T$ measured by \emph{Helios} in fast solar wind reveal \citep{hellinger11}
\begin{equation}\label{TpHelli}
T(r)=2.5\times 10^5\,\mathrm K \left(\frac{r}{1\,\mathrm{au}}\right)^{-0.74}.
\end{equation}
Like in the case of our \emph{Ulysses} measurements, the \emph{Helios} proton temperature profiles also depend on the solar-wind speed \citep{marsch82}. The power index for the scaling of the perpendicular proton temperature varies from $-1.17$ in fast wind to $-0.9$ in slow wind, and the power index for the scaling of the parallel proton temperature varies from $-0.69$ in fast wind to $-1.03$ in slow wind.  We combine the scaling in equation~(\ref{Tp}) with the scaling
\begin{equation}\label{TpH}
T=\tilde T\left(\frac{r}{1\,\mathrm{au}}\right)^{-0.74}
\end{equation}
according to equation~(\ref{TpHelli}) to estimate the $r$-dependence of the proton temperature from the location of the \emph{Ulysses} measurement to the heliocentric distance $r_{\mathrm S}$. We acknowledge that the assumption of a single power index neglects the differences in the temperature profiles between fast and slow wind and should be seen as an average estimate.

\begin{figure}
\includegraphics[width=\columnwidth]{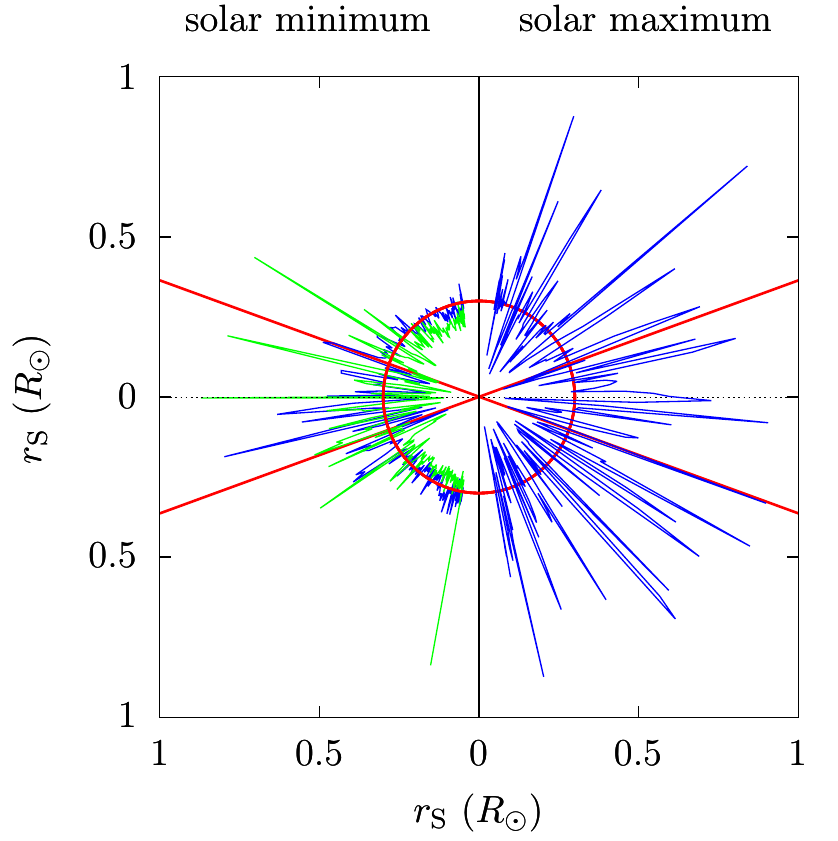}
    \caption{Polar plot of the estimated sonic radius $r_{\mathrm S}$ based on the scaled proton temperature during solar minimum (left half, FLS1 in blue, FLS3 in green) and solar maximum (right half, FLS2). The values of $r_{\mathrm S}$ give lower estimates for the sonic radius. We only plot $r_{
\mathrm S}$ if $r_{\mathrm S}>0$. The format of this plot is the same as in Figure~\ref{fig:dial_massflux}. The red circle has a radius of $0.3R_{\odot}$.}
  \label{fig:dial_rS}
\end{figure}

The procedure for the calculation of $r_{\mathrm S}$ is as follows. We first calculate $\tilde T$ according to equation~(\ref{Tp}). Then we apply the scaling in equation~(\ref{TpH}) to achieve an $r$-dependent value of $T(r)$ in the inner heliosphere. Based on this value, we calculate $c_{\mathrm S}(r)$ and evaluate the distance at which equation~(\ref{soniccond}) is fulfilled. This leads to
\begin{equation}\label{rS}
r_{\mathrm S}= (1\,\mathrm{au})\, \tilde M_{\mathrm S}^{-2.70}.
\end{equation}
As for our estimate of $r_{\mathrm A}$, equation~(\ref{rS}) gives a lower limit on the sonic radius due to our assumption that $\partial U_r/\partial r=0$ at $r\gtrsim r_{\mathrm S}$. We show the polar plot of the estimated sonic radius $r_{\mathrm S}$ according to equation~(\ref{rS}) in Figure~\ref{fig:dial_rS}. Overall, we observe an increase of $r_{\mathrm S}$ with increasing $|\lambda|$ both during solar minimum and solar maximum. During solar minimum, $r_{\mathrm S}$ varies between values of about $0.09R_{\odot}$ and $0.9R_{\odot}$ at equatorial heliolatitudes.  At polar heliolatitudes, $r_{\mathrm S}\approx 0.3R_{\odot}$ during both FLS1 and FLS3, although the average $r_{\mathrm S}$ is slightly smaller during FLS3. During solar maximum, the value of $r_{\mathrm S}$ varies between about $0.08R_{\odot}$ and $1.1R_{\odot}$.  We discuss in Section~\ref{sect:disc} that our assumptions in the derivation of equation~(\ref{rS}) are strongly violated at these distances, casting doubt on the reliability of our estimate of $r_{\mathrm S}$. In addition, numerical values of $r_{\mathrm S}<1R_{\odot}$, like for the case of $r_{\mathrm A}$, are clearly unphysical as they lie within the sphere of the Sun.

\begin{figure}
\includegraphics[width=\columnwidth]{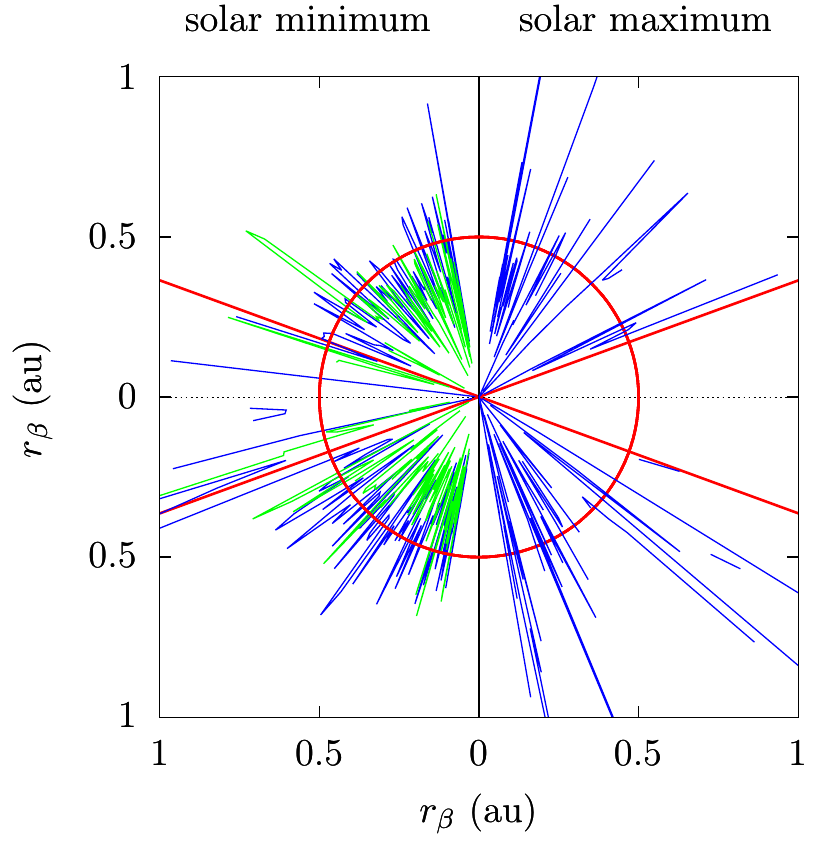}
    \caption{Polar plot of the estimated $\beta=1$ radius $r_{\beta}$ based on the scaled moment profiles during solar minimum (left half, FLS1 in blue, FLS3 in green) and solar maximum (right half, FLS2).  We only plot $r_{\beta}$ if $r_{\beta}>0$. The format of this plot is the same as in Figure~\ref{fig:dial_massflux}. The red circle has a radius of $0.5\,\mathrm{au}$.}
  \label{fig:dial_rBeta}
\end{figure}

Lastly, we estimate the radius $r_{\beta}$ at which $\beta=1$, where 
\begin{equation}
\beta=\frac{8\pi Nk_{\mathrm B}T}{B^2}
\end{equation}
is the ratio between the thermal pressure of the protons and the magnetic energy density. Since $r_{\beta}$ can be greater than or less than 1\,au, we must account for the different $T$-scalings according to equations~(\ref{Tp}) and (\ref{TpH}) in the outer and inner heliosphere.
The radius $r_{\beta}$ then fulfils the conditions 
\begin{equation}\label{rb1}
B_r^2+\left(\frac{r_{\beta}}{r}\right)^{2} B_{\phi}^2-8\pi N k_{\mathrm B}T\left(\frac{r_{\beta}}{r}\right)^{0.98}=0
\end{equation}
if $r_{\beta}>1\,\mathrm{au}$ and
\begin{equation}\label{rb2}
B_r^2+\left(\frac{r_{\beta}}{r}\right)^{2}B_{\phi}^2  -8\pi N k_{\mathrm B}T\left(\frac{r_{\beta}}{r}\right)^{1.26}\left(\frac{r}{1\,\mathrm{au}}\right)^{0.28}=0
\end{equation}
if $r_{\beta}<1\,\mathrm{au}$. We solve this set of conditional equations (\ref{rb1}) and (\ref{rb2}) numerically for $r_{\beta}$ through a Newton-secant method. We ignore all cases in which the solution leads to $r_{\beta}<0$. We show the polar plot of the estimated $\beta=1$ radius $r_{\beta}$ according to equations~(\ref{rb1}) and (\ref{rb2}) in Figure~\ref{fig:dial_rBeta}. The data gaps in Figure~\ref{fig:dial_rBeta} show that our method at times fails to provide a reliable estimate for $r_{\beta}$ in equatorial regions. Under solar-minimum conditions, we find solutions with $0.01\,\mathrm{au}<r_{\beta}<3.2\,\mathrm{au}$ in the equatorial plane for some time intervals. We find that $r_{\beta}$ assumes average values between about 0.2\,au and 0.5\,au at heliolatitudes polewards of $\pm 20^{\circ}$ during solar minimum. During FLS3, $r_{\beta}$ is on average less than $r_{\beta}$ during FLS1. During solar maximum, $r_{\beta}$ varies from 0 to about 4\,au. Like in the solar-minimum case, our approach does not always provide us with reliable estimates for $r_{\beta}$ near the equatorial plane during solar maximum. We find equatorial solutions with $0.2\,\mathrm{au}<r_{\beta}<2.6\,\mathrm{au}$ during maximum conditions.

\section{Discussion}\label{sect:disc}

Our measurement of the mass flux per steradian $\mathcal F_m$ in the ecliptic plane is consistent with earlier measurements \citep{feldman78,mccomas00,cranmer17,finley18}. These observations report that $4\pi r^2U_rNM\approx 10^{12}\,\mathrm{g/s}$, which corresponds to $\mathcal F_m\approx 4\times10^{-16}\,\mathrm{au}^2\,\mathrm{g\,cm}^{-2}\,\mathrm s^{-1}\,\mathrm{sr}^{-1}$ in our units. In agreement with previous measurements from \emph{Ulysses} during solar minimum  \citep{forsyth96,goldstein96}, we find that $\mathcal F_m$ is approximately constant for $|\lambda|>20^{\circ}$, although we find a slight decrease of $\mathcal F_m$ towards the poles during FLS1, especially during the Northern pass at solar minimum \citep[see also][]{mccomas00}. In general,  $\mathcal F_m$ is by almost 25\% smaller during the minimum recorded in FLS3 than during the minimum recorded in FLS1. This finding confirms the extraordinary conditions during the very deep solar minimum that extended into 2009  \citep{mccomas13}. The mass flux is a useful quantity to distinguish solar-wind acceleration models. As pointed out by \citet{holzer81}, for instance, a constancy of the mass flux with solar-wind speed would suggest that most of the energy deposition occurs above the Alfv\'en-critical point. The observed variability in $\mathcal F_m$, however, confirms the notion that most of the energy deposition occurs below the critical point. We note that $\alpha$-particles, which we neglect, can make a substantial contribution ($\lesssim 20\%$) to the solar-wind mass flux, especially in fast solar wind \citep{marsch82b}.

Our estimate for the momentum flux per steradian $\mathcal F_p$ is not fully scalable to different $r$ across the heliosphere, since it does not include the contributions from plasma-pressure gradients, magnetic-pressure gradients, and gravity. Although these effects are small at the location of the measurement, we warn that the scaling of $\mathcal F_p$ to the inner heliosphere deserves particular attention due to our application of the coasting approximation. Including $\alpha$-particles, which can make a substantial contribution to the momentum flux of the solar wind, \citet{mccomas00} report a momentum flux of approximately $2.9\times 10^{-8}\,\mathrm{au}^2\,\mathrm{g\,cm}^{-1}\,\mathrm s^{-2}\,\mathrm{sr}^{-1}$ during FLS1 in solar-minimum conditions with a variation of 0.3\% per degree in heliolatitude. Our measurements confirm that, both during solar minimum and (to a lesser degree) during solar maximum, the average momentum flux is lower in equatorial regions with only a small variation with heliolatitude in the polar regions \citep{phillips96}. We note, however, that, during all FLS orbits, the \emph{Ulysses} spacecraft has a larger latitudinal angular speed in the equatorial regions than in the polar regions. Therefore, the spacecraft may statistically encounter fewer transient events near the equator which potentially skews the measured variability.
The deep solar minimum during FLS3 exhibits polar $\mathcal F_p$-values that are by about 30\% smaller than the average polar $\mathcal F_p$ during FLS1. The momentum flux is of particular interest since it represents the solar-wind ram pressure, the key internal driver that determines the shape and extent of the heliosphere at the location of the heliospheric termination shock \citep{mccomas00,jokipii13}.

An earlier measurement of the solar wind's kinetic-energy density based on a combination of data from the \emph{Helios}, \emph{Ulysses}, and \emph{Wind} missions shows that $W=NMU_rU_r^2/2+NMU_rGM_{\odot}/R_{\odot}$ is largely independent of solar-wind speed at a value of approximately $1.5\times 10^{-3}\,\mathrm{W\,m}^{-2}$ at $r=1\,\mathrm{au}$ and in the plane of the ecliptic \citep{lechat12}. We note that $W$  depends on $r$, unlike our $\mathcal F_E$, which also includes contributions from the plasma pressure, and the magnetic field. Nevertheless, since these measurements are reported at $r=1\,\mathrm{au}$, we can compare $W$ with $\mathcal F_E$ by calculating $r^2W\approx 1.5\,\mathrm{au}^2\,\mathrm{g\,s}^{-3}\,\mathrm{sr}^{-1}$, which is slightly larger than our reported value. The estimate by \citet{lechat12} includes the required energy for the plasma to leave the Sun's gravitational potential from a distance of $r=1\,R_{\odot}$. Since even for fast wind $GM_{\odot}/R_{\odot}\sim U_r^2/2$, this additional contribution to $W$ explains the difference in these two estimates. The contribution of the magnetic field to $\mathcal F_E$ is small in agreement with modelling results \citep{alexander95}. At the distance at which \emph{Ulysses} measured $\mathcal F_E$, the energy flux is dominated by the kinetic-energy flux.  An earlier analysis of \emph{Ulysses} data reveals that the equatorial kinetic-energy flux is approximately 30\% less than the polar kinetic-energy flux  \citep{phillips96} consistent with our results, although this earlier study did not analyse its heliolatitudinal dependence during conditions near solar maximum. The deep solar minimum during FLS3 exhibits a polar energy flux that is by about 34\% smaller than the polar energy flux during FLS1.

Due to pointing uncertainties explicitly communicated by the instrument team, \emph{Ulysses} is unable to provide us with a reliable measurement of $U_{\phi}$ on the required time-scales. $U_{\phi}$ is a key component in the calculation of $\mathcal F_L$. Therefore, our polar plot of $\mathcal F_L$ in Figure~\ref{fig:dial_angmomflux} must be treated with caution. In addition, $U_{\phi}$ often exhibits relative variations greater than order unity so that it even assumes negative values in the solar wind. Natural variations due to turbulence or stream-interaction regions cause significant deflections compared to the expected average \citep{egidi69,siscoe69}. Although time averaging reduces the impact of these natural variations on the determination of the average $U_{\phi}$, the instrumental shortcomings remain. 
In addition to uncertainties in $U_{\phi}$, the accuracy of $\mathcal F_L$ also depends on the average values of $B_r$ and $B_{\phi}$. Although \emph{Ulysses} observations show that the average interplanetary magnetic field largely agrees with Parker's prediction, $B_r$ decreases more slowly and $B_{\phi}$ more quickly with heliocentric distance than expected \citep{mccomas00}. These deviations are usually attributed to waves or more complex boundary conditions near the Sun than those assumed in Parker's model \citep{forsyth96,forsyth96b}.  In Section~\ref{scalings}, we scale the magnetic field towards smaller $r$ assuming the radial scaling according to Parker's model. We expect that our averaging over 30 h removes most of the Alfv\'enic fluctuations that lead to variations in  $U_{\phi}$ and $B_{\phi}$ around their mean values. Some of the inaccuracies due to complex boundary conditions can be reduced using a more complex model for the interplanetary magnetic field such as the \citet{fisk96} model, which accounts for the Sun's differential rotation and the tilt of the Sun's magnetic axis compared to its rotational axis. However, such a more complex treatment is beyond the scope of this work. 

Earlier measurements of the Sun's angular-momentum loss provide a value of $\mathcal F_L\approx 5\dots 10\times 10^{-10}\,\mathrm{au}^3\,\mathrm{g\,cm}^{-1}\,\mathrm s^{-2}\,\mathrm{sr}^{-1}$ \citep{pizzo83,finley18,finley19,verscharen21}. Despite the inaccuracies in the \emph{Ulysses} measurement, these values are consistent with our estimates of $\mathcal F_L$ in the southern hemisphere during FLS1, in the Northern hemisphere during FLS3, and during solar maximum (FLS2). The contribution to the angular-momentum flux from $\alpha$-particles, which we neglect in our estimate, can be significant in the solar wind \citep{pizzo83,marsch84,verscharen15} and should be included in future studies with spacecraft that provide readily available $\alpha$-particle data. The partition between the particle contributions and the magnetic-field contributions to $\mathcal F_L$ is an important question with implications for solar-wind models. In the early \citet{weber67} model,  the field contribution is greater than the particle contribution; however, later measurements show that the partition can be opposite, especially in slow wind \citep{hundhausen70b,marsch84}. In typical fast wind, most of the angular-momentum is lost through magnetic stresses, an effect which is expected to be even stronger in the inner heliosphere \citep[see also][]{pizzo83,alexander95}. Likewise, the on average larger pressure anisotropies in the inner heliosphere \citep{marsch82} contribute to the angular-momentum equation \citep{hundhausen70}, even though we neglect their effect in our treatment based on the assumption of a scalar pressure. Due to the limitations in our measurement of $U_{\phi}$, we are unable to determine reliably the partition of these different contributions to the overall angular-momentum loss. In the very inner heliosphere, \emph{Parker Solar Probe} reports an azimuthal  flow which is significantly larger than expected by the \citet{weber67} model \citep{kasper19}. The source of this large $U_{\phi}$ component is still unclear. A stronger effective co-rotation or deflections through, for instance, stream interactions serve as potential explanations \citep{finley20}. In this context, a more complex field geometry near the Sun also affects the angular-momentum loss \citep{finley17}, which introduces further dependencies on the solar cycle \citep{reville17,finley18}.

A combination of \emph{Ulysses} measurements with a self-similar solar-wind model finds a scaled Alfv\'en Mach number $\tilde M_{\mathrm A}$ of about 10 to 20 \citep{sauty05} consistent with our simpler scaling estimate in Section~\ref{sect:sonic}. The earlier self-similar model, however, suffered from inconsistencies in reproducing observed magnetic-field values at 1\,au, which could be resolved in a later update \citep{aibeo07}. In the deeper solar minimum during FLS3, the polar $\tilde M_{\mathrm A}$ is on average slightly greater than the polar $\tilde M_{\mathrm A}$ measured during FLS1.

Our estimates of the critical radii $r_{\mathrm A}$, $r_{\mathrm S}$, and $r_{\beta}$ assume a constant radial bulk speed of the solar wind between the location of measurement and the respective critical radius. In the inner heliosphere, this assumption can be violated, especially regarding the slow solar wind \citep{schwenn81}. Therefore, our calculations only provide lower estimates for the location of these critical radii. The location of the Alfv\'en radius depends on the magnetic-field geometry near the Sun, which deviates from our simplifying  assumption of a Parker profile \citep{finley17}. A dipolar solar braking model constrained by \emph{Ulysses} data estimates that $r_{\mathrm A}\approx 16R_{\odot}$, independent of heliolatitude \citep{li99}. This value is consistent with our largest estimate in regions outside equatorial heliolatitudes. We find a larger variation between this model estimate and ours though. Our findings are in agreement with earlier estimates based on \emph{Helios} measurements in the ecliptic plane, which suggest that $r_{\mathrm A}\approx 12\dots 17R_{\odot}$ \citep{pizzo83,marsch84}.
Previous scalings based on a hydrodynamic model also estimate the values of $r_{\mathrm A}$ and $r_{\mathrm S}$ \citep{exarhos00}. During solar-minimum conditions, these models suggest that $r_{\mathrm A}\approx 14R_{\odot}$ and $r_{\mathrm S}\approx 1.5R_{\odot}$ at polar heliolatitudes, and that $r_{\mathrm A}\approx 17R_{\odot}$ and $r_{\mathrm S}\approx 2R_{\odot}$ in equatorial regions \citep[for an extension of this model, see also][]{katsikas10}. While our estimates for $r_{\mathrm A}$ are largely consistent with this hydrodynamic model within the observed variability, our estimates of $r_{\mathrm S}$ are smaller by a factor of approximately four to five. 
The reason for this discrepancy in our model is based on the complication that the condition $U_r=c_{\mathrm S}$ is often fulfilled in a region where significant solar-wind acceleration is still ongoing.
Since typically $r_{\mathrm S}<r_{\mathrm A}$, our estimate of $r_{\mathrm S}$ suffers  more strongly from the violation of our assumption that $\partial U_r/\partial r=0$ than our estimate of $r_{\mathrm A}$. In addition, the extrapolation of the $T$-profile according to equation~(\ref{TpHelli}) from $r=0.3\,\mathrm{au}$ to distances of a few $R_{\odot}$ is highly problematic. These shortcomings can be overcome in the future by using (i) a more realistic acceleration profile near the Sun and (ii) a more realistic $T$-profile near the Sun based on, for example, measurements from \emph{Parker Solar Probe}.
Based on OMNI data and their hydrodynamic model, \citet{exarhos00} predict a dependence of $r_{\mathrm A}$ on the solar cycle, which our measurements clearly confirm \citep[see also][]{kasper19a}. However, we cannot confirm two predictions made by this model: (i) a more spherical shape of both critical surfaces during solar maximum and (ii) a small variation of $r_{\mathrm S}$ with solar cycle.
Our estimated positions of $r_{\mathrm A}$ and $r_{\beta}$ are largely consistent with predictions from a magnetohydrodynamics model of the solar wind over polar regions \citep{chhiber19}. We note, however, that our finding of a larger value of $r_{\beta}\sim 1\,\mathrm{au}$ in the ecliptic plane is more consistent with the observation that, on average, $\beta\sim 1$ at the first Lagrange point \citep{wilson18}. Under the deep solar-minimum conditions during FLS3, we find that, over the poles, the positions of $r_{\mathrm A}$, $r_{\mathrm S}$, and $r_{\beta}$ are on average slightly closer to the Sun than during FLS1.

Our calculation neglects any super-radial expansion effects due to expanding flux tubes in the solar wind. These effects can be significant at very small distances from the Sun. Above a few solar radii, however, the super-radial expansion of the coronal magnetic field is expected to be small \citep{woo97}, although this expectation has been discussed controversially \citep{neugebauer99}. If important, the super-radial expansion would especially affect our calculations of $r_{\mathrm A}$ and $r_{\mathrm S}$. Nevertheless, all higher-order multipoles of the coronal magnetic field eventually (probably beyond a few solar radii) drop faster than the dipole moment, so that super-radial expansion then becomes negligible \citep{sandbaek94,wang97}.

\section{Conclusions}

We use proton and magnetic-field data from the \emph{Ulysses} mission to study the dependence of mass, momentum, energy, and angular-momentum fluxes on heliolatitude. Based on the multifluid framework and assuming an isotropic electron--proton plasma, we derive laws for the radial conservation of these fluxes. These conservation laws allow us to separate the radial dependence from the heliolatitudinal dependence in the \emph{Ulysses} measurements. A major caveat of this method lies in the neglect of the natural spatio-temporal variations in the solar-wind plasma and magnetic field which occur over a wide range of scales \citep{verscharen19}. Therefore, our analysis only applies to the average large-scale behaviour of the solar wind. Moreover, we neglect effects due to temporal and heliolongitudinal changes in the source regions of the solar wind. The variability of the flux parameters shown in our analysis, especially during solar maximum, give us an estimate for the natural variability of the solar wind on time-scales greater than our averaging time of 30\,h.

Although the \emph{Ulysses} data set is unprecedented in its heliolatitudinal coverage of the solar wind, we expect major advances regarding the topics addressed in this work from the ongoing measurements from \emph{Parker Solar Probe} and \emph{Solar Orbiter}. 
 \emph{Parker Solar Probe} will explore the very inner regions of the heliosphere. Our analysis suggests that it will cross the distance $r_{\mathrm A}$ during the later stages of its orbit, while it is unlikely to cross the distance $r_{\mathrm S}$.  \emph{Solar Orbiter} will leave the plane of the ecliptic during its extended mission phase and measure the solar wind at heliolatitudes up to $\pm 33^{\circ}$. Considering that the axial tilt of the Sun's magnetic-field dipole axis is $\lesssim 10^{\circ}$ during solar minimum \citep{norton08}, \emph{Solar Orbiter} will cover an even larger range of heliomagnetic latitudes during that phase of the mission\footnote{During the early phase of the \emph{Ulysses} mission, a tilt between the Sun's magnetic-field dipole axis and its rotational axis of about $30^{\circ}$ has been reported \citep{bame93,hoeksema95}.}.  These measurements will, therefore, allow us to study the heliolatitudinal dependence of the relevant solar-wind parameters in a similar way to this study. In addition, both \emph{Parker Solar Probe} and \emph{Solar Orbiter} also provide us with high-resolution measurements of the electron distribution function. Although the electron contributions to the mass, momentum, and angular-momentum fluxes are negligible, their contribution to the energy flux is significant in the form of heat flux \citep{ogilvie71,hollweg74,feldman75}. It will be worthwhile to include this effect in future studies of this kind. These new observations will help us to further constrain solar-wind models and drive forward our understanding of the acceleration of the solar wind.

Lastly, we emphasize that our observations support the general picture that the solar wind is much more variable during times of solar maximum than during times of solar minimum. This increased variability is likely to be caused by transient events such as interplanetary coronal mass ejections (ICMEs). In our analysis, we include such events in order to reflect the range and pattern of the overall variability on the investigated time-scales. However, it would be worthwhile in a future study to separate our data set into time intervals with and without ICMEs according to existing ICME catalogues for the \emph{Ulysses} data set \citep[e.g.,][]{ebert09,du10,richardson14}. This approach would facilitate a detailed study of the contribution of transient events to the variability of mass, momentum, energy, and angular-momentum fluxes. We note, however, that some of our model assumptions, such as the azimuthal symmetry, non-polar field and flow components, and radial scalings must be treated with caution in transient events. 
In addition, we also find a significant variation in some of the analysed quantities between the solar minima recorded during FLS1 and FLS3. It is of interest to study these conditions more closely. This should especially include a closer inspection of the question to what degree the observed variations are real and to what degree they result from breakdowns in our assumptions.

\section*{Acknowledgements}

DV~is supported by the Science and Technology Facilities Council (STFC) Ernest Rutherford Fellowship ST/P003826/1 and STFC Consolidated Grant ST/S000240/1. SDB~acknowledges the support of the Leverhulme Trust Visiting Professorship programme.  The authors acknowledge insightful discussions within the International Team ``Exploring The Solar Wind In Regions Closer Than Ever Observed Before'' at the International Space Science Institute (ISSI) in Bern led by Louise Harra. We acknowledge the National Space Science Data Center for the provision of the \emph{Ulysses} data.

\section*{Data availability}

All used data are freely available online at NASA's National Space Science Data Center under \url{https://nssdc.gsfc.nasa.gov}.

%%%%%%%%%%%%%%%%%%%%%%%%%%%%%%%%%%%%%%%%%%%%%%%%%%



\bibliographystyle{mnras}
\bibliography{dial_plots_rev} % if your bibtex file is called example.bib

%%%%%%%%%%%%%%%%%%%%%%%%%%%%%%%%%%%%%%%%%%%%%%%%%%

%%%%%%%%%%%%%%%%% APPENDICES %%%%%%%%%%%%%%%%%%%%%

%%%%%%%%%%%%%%%%%%%%%%%%%%%%%%%%%%%%%%%%%%%%%%%%%%

% Don't change these lines
\bsp	% typesetting comment
\label{lastpage}
\end{document}